\documentclass[aps,pra,twocolumn,superscriptaddress,longbibliography]{revtex4-2}

\pdfoutput=1
\usepackage[english]{babel}
\usepackage{xcolor}
\newcommand{\comment}[1]{}
\usepackage{amsmath}
\usepackage{amsfonts}
\usepackage{tensor}
\usepackage{physics}
\usepackage{graphicx}
\usepackage[colorlinks=true, allcolors=blue]{hyperref}
\usepackage{siunitx}

\begin{document}

\title{Optimal recoil-free state preparation in an optical atom tweezer}
\author{Lia Kley}
\email{lkley@physnet.uni-hamburg.de}
\affiliation{Zentrum f\"ur Optische Quantentechnologien, Universit\"at Hamburg, 22761 Hamburg, Germany}
\affiliation{Institut für Quantenphysik, Universit\"at Hamburg, 22761 Hamburg, Germany}
\author{Nicolas Heimann}
\affiliation{Zentrum f\"ur Optische Quantentechnologien, Universit\"at Hamburg, 22761 Hamburg, Germany}
\affiliation{Institut für Quantenphysik, Universit\"at Hamburg, 22761 Hamburg, Germany}
\affiliation{The Hamburg Centre for Ultrafast Imaging, 22761 Hamburg, Germany}
\author{Aslam Parvej}
\affiliation{Zentrum f\"ur Optische Quantentechnologien, Universit\"at Hamburg, 22761 Hamburg, Germany}
\affiliation{Institut für Quantenphysik, Universit\"at Hamburg, 22761 Hamburg, Germany}
\author{Lukas Broers}
\affiliation{Zentrum f\"ur Optische Quantentechnologien, Universit\"at Hamburg, 22761 Hamburg, Germany}
\affiliation{Institut für Quantenphysik, Universit\"at Hamburg, 22761 Hamburg, Germany}
\author{Ludwig Mathey}
\affiliation{Zentrum f\"ur Optische Quantentechnologien, Universit\"at Hamburg, 22761 Hamburg, Germany}
\affiliation{Institut für Quantenphysik, Universit\"at Hamburg, 22761 Hamburg, Germany}
\affiliation{The Hamburg Centre for Ultrafast Imaging, 22761 Hamburg, Germany}

\begin{abstract}
Quantum computing in atom tweezers requires high-fidelity implementations of quantum operations. Here, we demonstrate the optimal implementation of the transition $|0\rangle \rightarrow |1\rangle$ of two levels, serving as a qubit, of an atom in a tweezer potential,  driven by a single-photon Rabi pulse. The Rabi pulse generates a photon recoil of the atom, due to the Lamb-Dicke coupling between the internal and motional degree of freedom, driving the system out of the logical subspace. This detrimental effect is strongly suppressed in the protocols that we propose. Using pulse engineering, we generate optimal protocols composed of a Rabi protocol and a force protocol, corresponding to dynamically displacing the tweezer. We generate these for a large parameter space, from small to large values of the Rabi frequency, and a range of pulse lengths. We identify three main regimes for the optimal protocols, and discuss their properties. In all of these regimes, we demonstrate infidelity well below the current technological standard, thus mitigating a universal challenge in atom tweezers and other quantum technology platforms.
\end{abstract}
\maketitle

\section{Introduction}
\label{sec:introduction}
The identification and development of a platform for quantum computing constitutes a profound intellectual and technological challenge. Competing requirements have to be fulfilled. On the one hand, fast and precise operations have to be implemented via external manipulation or via strong interactions. On the  other hand, any undesired interactions or externally induced dynamics have to be suppressed, given the demand for the large number of high-fidelity operations that a quantum computer has to support. A platform under rapid development are atom tweezer arrays~\cite{Saffman_2016, PhysRevLett.87.037901, morgado_quantum_2021, Browaeys2020Nature-Physics, Graham2022-Nature, PRXQuantum.2.030322}. In these systems, the atoms are held individually in optical tweezer traps, and controlled via external laser pulses and brought to interaction via the strong van-der-Waals interactions that are present in Rydberg states. Given the finite magnitude of the optical trap potential, it is imperative that the atoms are prepared in the motional ground state of the trap, e.g. via Raman sideband cooling~\cite{PhysRevLett.110.133001, PhysRevA.97.063423, PhysRevResearch.5.033093, Spence_2022}. Additionally, it is imperative that the atoms remain in the ground state during the operation of a quantum algorithm. Thus, the motional degree of freedom is a potential source of decoherence, resulting in an imperfect realization of a quantum computer. We note that recent studies on the interplay of motional and internal degrees of freedom in these systems have been reported in~\cite{PhysRevLett.127.050501,zhang2024recoil}.

\begin{figure}
    \centering
    \includegraphics[width=8.5cm]{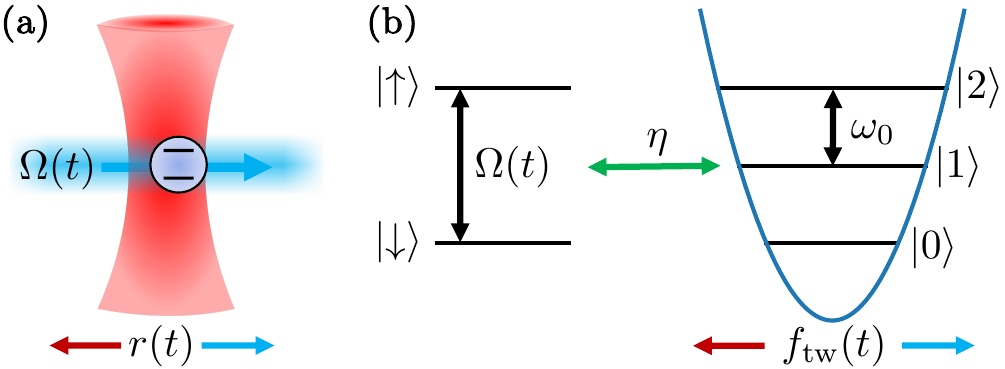}
    \caption{\textbf{Recoil-free state preparation.}
    \textbf{(a)}
    Sketch of an atom held in an optical tweezer (red) and driven by a control laser (blue) with Rabi frequency $\Omega(t)$. The photon recoil, indicated by the blue arrow, induces atomic motion. We propose to use the displacement of the trap $r(t)$ as part of the protocol to mitigate photon recoil. \textbf{(b)} The states $\ket{\downarrow}$ and $\ket{\uparrow}$ represent the atomic degree of freedom and the states $\ket{n}$ represent the motional degree of freedom of the atom in a trap with a trap frequency $\omega_0$. The system is driven on-resonance with a Rabi frequency $\Omega(t)$ and the coupling strength between the spin and the internal states is characterized by the Lamb-Dicke parameter $\eta$. Starting from the initial state $\ket{\downarrow,0}$, the objective is to prepare the state $\ket{\uparrow,0}$. The displacement $r(t)$ of the trap corresponds to a restoring force $f_\text{tw}(t)$. We propose optimal protocols based on $\Omega(t)$ and $f_\text{tw}(t)$.}
    \label{fig:1}
\end{figure}

Quantum optimal control is a powerful and versatile heuristic to optimize quantum dynamical processes according to a desired metric. Optimization algorithms are widely used in noisy intermediate scale quantum devices~\cite{Preskill2018quantumcomputingin,RevModPhys.94.015004} such as superconducting qubits~\cite{PhysRevB.79.060507, Egger_2014, PhysRevA.90.012318, Werninghaus2021-npj-Quantum-Information}, trapped ions \cite{PhysRevLett.112.190502, Figgatt2019-Nature, PhysRevApplied.16.024039}, and neutral atoms~\cite{PhysRevA.84.042315, PhysRevA.90.032329, PRXQuantum.4.030323, jandura_time-optimal_2022, heimann2023quantum} platforms.  Numerous optimization methodologies and algorithms have been put forth, including variational quantum algorithms and its extensions. A widely used method of pulse engineering is Gradient Ascent Pulse Engineering (GRAPE)~\cite{khaneja2005optimal}, which infers the optimal transformation parameters in an iterative fashion based on gradient ascent such that gradients are approximated by finite differences. This method can be extended to Pulse Engineering via Projection of Response Functions (PEPR)~\cite{heimann_2024-PEPR} by utilizing linear response theory to estimate gradients.

In this paper we present a recoil-free implementation of the transition $\ket{\downarrow}\rightarrow \ket{\uparrow}$ of two internal states $\ket{\downarrow}$ and $\ket{\uparrow}$ of an atom held in an optical tweezer. While for an atom held in an infinitely strong trap, this is achieved via a standard Rabi $\pi$-pulse, the photon recoil of the laser pulse activates the atomic motion in a finite trap. Therefore, the recoil-free protocols that we present here, control the atomic state to be preserved in the logical subspace and in the motional ground state. We propose to use both the Rabi amplitude and phase, as well as the tweezer position as part of the protocol. As the central result, we develop an optimal protocol via the pulse engineering method PEPR. As we discuss, we achieve very low infidelities for a wide range of parameters. We identify three regimes, slow, intermediate and fast, which are primarily determined by the ratio of the maximal Rabi frequency and the trap frequency. Each of the regimes is characterized by low infidelity, in a plateau-like fashion. We compare this optimal protocol with a recoil-compensated Rabi protocol that we propose. This protocol approximates the performance of the optimal protocol in some of the identified regimes, and can therefore be applied in experiment deterministically.  Furthermore, we discuss how the infidelity of these protocols is modified in the presence of dissipation. While dissipation introduces a lower bound on the infidelity, we again find a very low infidelity of the optimal protocols, demonstrating the robustness of our approach.

This paper is structured as follows. In Sect.~\ref{sec:model}, we introduce and motivate the underlying model of this manuscript. In Sect.~\ref{sect:optimal-strategies}, we outline strategies to determine optimal control protocols. In Sect.~\ref{sec:high-fidelity-operations}, we demonstrate recoil-free state preparation with and without dissipation. In Sect.~\ref{sec:conclusion}, we conclude our findings.

\section{Single atom in an optical tweezer}
\label{sec:model}
We consider a single atom held in an optical tweezer, as shown in Fig.~\ref{fig:1}~(a). We include two atomic states in our model, which we refer to as $\ket{\downarrow}$ and $\ket{\uparrow}$. In addition to these internal states of the atom, we include the motional states of the atom in the optical trap potential. We approximate this as a harmonic potential with frequency $\omega_0$, resulting in the Hamiltonian of the motional degree of freedom
\begin{equation}
    H_m = \hbar \omega_0 a^\dagger a.
\end{equation}
Thus, the basis states of our model are $\{\ket{\downarrow/\uparrow, n}\}$. We drive the transition between $\ket{\downarrow}$ and $\ket{\uparrow}$ resonantly with a Rabi frequency of $\Omega(t) = h_x(t)+ih_y(t)$, as depicted in Fig.~\ref{fig:1}~(a). $h_x(t)$ and $h_y(t)$ are the real valued amplitudes of the driving term. As a defining energy scale we introduce the maximal Rabi frequency $\Omega_\text{max}$ which we impose on the Rabi frequency, i.e. $|\Omega(t)|<\Omega_\text{max}$. For a perfectly confined atom, the Rabi driving would induce oscillations between the internal states $\ket{\uparrow}$ and $\ket{\downarrow}$ only. However, due to the finite confinement of the optical trap, the photon induced recoil of the Rabi driving induces a spatial motion as well. With this, the Rabi coupling takes the form
\begin{equation}
    H_\text{dr}(t) = \frac{\hbar\Omega(t)}{2}\sigma^+e^{i\eta(a+a^\dagger)} + \text{h.c.},
    \label{eq:H-dr-1}
\end{equation}
written in the rotating frame and the rotating wave approximation. $\eta=\sqrt{\hbar/(2m\omega_0)}k$ is the Lamb-Dicke parameter, $m$ is the atomic mass, $k$ is the wave vector of the control laser and $x_0$ is the harmonic oscillator length of the trap. We further consider the spatial displacement $r(t)$ of the trap potential of the tweezer, as shown in Fig.~\ref{fig:1}~(a). This displacement results in a Hamiltonian of the form
\begin{equation}
   H_\text{tw}(t) = -\hbar f_\text{tw}(t)(a+a^\dagger).
   \label{eq:Htw}
\end{equation}
Here $\hbar f_\text{tw}(t)~=~F(t)x_0$ is the scale of the restoring potential with the restoring force $F(t)~=~m\omega_0^2 r(t)$. $f_\text{tw}(t)~=~x_0F(t)/\hbar$ is the normalized force, with units of frequency. As a second defining energy scale, we impose a maximal velocity $v_\text{max}$ on the tweezer displacement, i.e. $|\partial_t r(t)| < v_\text{max}$. The resulting model
\begin{equation}
    H = H_m  + H_\text{dr}(t) + H_\text{tw}(t),
    \label{eq:H-tot}
\end{equation}
is depicted schematically in Fig.~\ref{fig:1}~(b).

As an example for dissipation, we include dephasing between the states $\ket{\uparrow}$ and $\ket{\downarrow}$. We include this as a Lindblad operator $\sigma_z$ in a master equation, with the dephasing rate $\gamma_z$. This gives the equation of motion
\begin{equation}
    \frac{d{\rho}}{dt} = -\frac{i}{\hbar}[H,\rho] + \gamma_z(L \rho L^\dagger -\rho).
    \label{eq:drho}
\end{equation}
where $\rho$ is the density matrix of the full system, $L~=~\sigma_z\otimes 1_m$ with the identity $1_m$  on the motional subspace.

As we describe below, the ratio of the maximal Rabi frequency $\Omega_\text{max}$ and the trap frequency $\omega_0$ determines qualitatively distinct dynamical regimes. We refer to the regime of $\Omega_\text{max} \gg \omega_0$ as the fast regime, to the regime of $\Omega_\text{max} \ll \omega_0$ as the slow regime, and to the regime with $\Omega_\text{max} \approx \omega_0$ as the intermediate regime.

As our main example for the fast regime, we consider $~^{171}$Yb atoms trapped in an optical tweezer potential with a trap frequency of $\omega_0/(2\pi)=50\si{\kilo\hertz}$~\cite{chen_analyzing_2022}. Specifically, we consider the single-photon Rabi coupling driven by an ultraviolet control laser, i.e. $k=2\pi/(302\si{\nano\meter})$, between the $\ket{\downarrow}=\ket{^3P_0, m_F=-1/2}$ state and a highly excited Rydberg state $\ket{\uparrow}=\ket{^3S_1}$. In this case, the Lamb-Dicke parameter is $\eta=0.505$. As a typical range for the Rabi frequency, we consider $\Omega_\text{max}=1-10\si{\mega\hertz}$. This is large compared to the trap frequency, i.e. $\Omega_\text{max} \gg \omega_0$, and therefore in the fast regime. As an example for the slow regime, we consider the single-photon Rabi coupling of the optical clock transition between $\ket{\downarrow}=\ket{^1S_0, m_F=-1/2}$ and $\ket{\uparrow}=\ket{^3P_0, m_F=1/2}$, driven by a microwave laser. In this regime, the maximal Rabi frequency $\Omega_\text{max}$ is on the order of $\si{\kilo\hertz}$. Here, a typical value of the Lamb-Dicke parameter is $\eta=0.2158$. However, to simplify the analysis in this work we use $\eta=0.505$ in both regimes and we emphasize that our model and the numerical optimization applies to any atomic species and optical potential.

\section{Optimal control strategies}
\label{sect:optimal-strategies}
Our objective is to identify an optimal strategy to generate the transformation of $\ket{\downarrow,0}$ to $\ket{\uparrow,0}$. In the absence of Lamb-Dicke coupling to the motional degree of freedom, i.e. for $\eta = 0$, this transformation is generated by a $\pi$-pulse in the atomic sector. However, for $\eta > 0$, the laser-induced dynamics generates population of the states $\ket{\uparrow/\downarrow, n > 0}$, i.e. of states outside of the desired logical subspace. Therefore, the objective of driving the initial state $\ket{\downarrow,0}$ into the target state $\ket{\uparrow,0}$ corresponds to suppressing the excitation of the motion of the atom, and of operating exclusively within the logical subspace. We consider the initial state $\rho_0=\ket{\downarrow,0}\bra{\downarrow,0}$ in a density matrix representation, and refer to the numerically time-propagated state over the time interval $[0, t]$, according to Eq.~\ref{eq:drho}, as $\rho(t)$. We truncate the motional state space with a maximal value $n_\text{max}=3$, with $n = 0, \dots , n_\text{max}$. The target state of the transformation is $\rho_*=\ket{\uparrow,0}\bra{\uparrow,0}$. To generate an optimal strategy, we minimize the infidelity of the state $\rho(t)$ compared to the target state
\begin{equation}
    1-F=1-\text{Tr}(\rho_*^\dagger \rho(t_f))
    \label{eq:1-F}
\end{equation}
where $t_f$ is the transformation time. 

\subsection{Rabi protocol}
\label{subsect:optimal-strategies-Rabiprot}
As we describe below, we generate the optimal protocol via the pulse engineering method PEPR. For comparison, we discuss two additional protocols, the Rabi protocol and the recoil-compensated Rabi protocol. For the Rabi protocol, we ignore the Lamb-Dicke coupling, and consider the pulse area
\begin{equation}
    \Theta(t)=\int_0^{t}d\tau |\Omega(\tau)|.
\end{equation}
We consider the functional form $\Omega(t)=\Omega_\text{max}\sin(\pi t / t_f)$. As we discuss below, we utilize the set of sine functions $\sin(\pi l t/t_f)$ for the numerical optimization. Thus, this choice of $\Omega(t)$ corresponds to the $l=1$ mode, and serves as a comparison. To generate a $\pi$-pulse, we use $\pi = \Theta(t_f)$ to determine the relation between the pulse length $t_f$ and $\Omega_\text{max}$, which is $t_f=\pi^2/(2\Omega_\text{max})$.  While we have motivated this relation between $t_f$ and $\Omega_\text{max}$ for the limit of $\eta = 0$, we apply the protocol to the case $\eta>0$ in the following, and refer to the generated infidelity as $1 - F_0$.

\subsection{Recoil-compensated Rabi protocol}
\label{subsect:optimal-strategies-RecoilcompRabiprot}
The second protocol refines the Rabi protocol by including both real and imaginary part of $\Omega(t)$ into the protocol, and by including a force protocol $f_\text{tw}(t)$ of the normalized force. To motivate the magnitude and time dependence of $f_\text{tw}(t)$, we note that during a $\pi$-pulse, the atom absorbs the momentum $\hbar k$. To compensate this momentum, the restoring force, see Eq.~\ref{eq:Htw}, has to fulfill:
\begin{equation}
    \hbar k + \int_0^{t_f} dt F(t) = 0.
    \label{eq:hk}
\end{equation}
Expressed via the Lamb-Dicke parameter $\eta=k x_0$ and the normalized force, we have
\begin{equation}
    \int_0^{t_f} dt f_\text{tw}(t) = -\eta.
    \label{eq:j}
\end{equation}
In the following, we denote the integral $\int dt f_\text{tw}(t)$ as the normalized impulse $j$. As we discuss below, the protocols that we obtain via numerical optimization in the fast regime, approximately display $j = -\eta$. In addition to the force protocol, we include an on-resonance Rabi driving, i.e. $\Omega(t)=h(t)e^{i\varphi}$, where $h(t)$ is the real-valued amplitude of the Rabi frequency and $\varphi$ is the phase. We expand the driving Hamiltonian Eq.~\ref{eq:H-dr-1} to first order in $\eta$
\begin{equation}
\begin{split}
    H_\text{dr}/\hbar &\approx \frac{h(t)}{2}(\cos(\varphi)\sigma_x-\sin(\varphi)\sigma_y) \\&- \eta\frac{h(t)}{2}(\sin(\varphi)\sigma_x + \cos(\varphi)\sigma_y)(a+a^\dagger).
    \label{eq:H-dr-firstorder}
\end{split}
\end{equation}
This motivates recoil compensation at first order in $\eta$ via the following force protocol
\begin{equation}
    f^{h}_\text{tw}(t) = \eta\frac{h(t)}{2}\langle \sin(\varphi)\sigma_x(t) + \cos(\varphi)\sigma_y(t) \rangle.
     \label{eq:ftw-recoil}
\end{equation}
The expectation value, at 0-th order in $\eta$, evaluates to
\begin{equation}
    \langle \sin(\varphi)\sigma_x(t) + \cos(\varphi)\sigma_y(t) \rangle = -\sin\left(\int_0^{t} d\tau h(\tau)\right),
    \label{eq:ev-expval}
\end{equation}
see App.~\ref{app:sec-derivation-exp}. This protocol induces a momentum of $-\hbar k$, such that it reduces the excitation of the motional states during the transformation, as we show in App.~\ref{app:sec-impulse}. As a concrete example, and in analogy to the Rabi protocol, we consider $\Omega(t)=\Omega_\text{max}\sin(\pi t/t_f)$ with $t_f=\pi^2/(2\Omega_\text{max})+\delta t$. $\delta t$ is a correction to the final time $t_f$, as we describe below. With this, the force protocol is
\begin{equation}
    f_\text{tw}^{\sin}(t) = -\frac{\eta \Omega_\text{max}}{2}\sin\left(\frac{\pi t}{t_f}\right)\sin\left(\pi\sin^2\left(\frac{\pi t}{2t_f}\right)\right).
    \label{eq:ftw}
\end{equation}
We refer to this protocol as the recoil-compensated Rabi protocol and we refer to the corresponding infidelity as $1-F_f$. We choose the correction to the pulse length $\delta t$ as follows. For $\Omega_\text{max}/\omega_0 \ll 1$, i.e. the limit of weak driving strength, the motional activation of the oscillator is small, thus the Rabi frequency is dominated by the ground state Debye-Waller factor. Specifically, the Rabi frequency is effectively reduced by the Debye-Waller factor of the ground state of the oscillator, i.e. $\Omega_\text{eff} = \langle 0| \exp(i\eta(a+a^\dagger)) | 0 \rangle = \Omega_\text{max} \exp(-\eta^2/2)$ . For $\Omega_\text{max}/\omega_0 \gg 1$, i.e. for strong driving, the motion of the atom is easily activated, and the atom responds to the photon recoil approximately freely. Therefore the two-level system is driven with the bare Rabi frequency, such as $\Omega_\text{max}$. As a simple approximation for this dynamical response, we choose $\delta t = \exp(\eta^2/2)-1$ for $\Omega_\text{max}/\omega_0<1$, and $\delta t = 0$ for  $\Omega_\text{max}/\omega_0>1$.

In addition to this protocol, we also introduce a force protocol that is derived from an arbitrary Rabi pulse $h(t) =|\Omega(t)|$. Based on $h(t)$ we define
\begin{equation}
f_{\text{tw},0}^{h} (t) = - \frac{\eta h(t)}{2} \sin\left(\int h(\tau)d\tau\right)
\end{equation}
motivated by Eq.~\ref{eq:ftw-recoil} and Eq.~\ref{eq:ev-expval}. We then extract the parameters of an expansion in sine functions via
\begin{equation}
\theta_{f,l}  = 2 \int_0^{t_f} dt  \sin\left(\pi l \frac{t}{t_f}\right) f_{\text{tw},0}^{h}(t) .
\end{equation}
We use these parameters to generate a force protocol $f_\text{tw}^{\text{pr}}(t)$. As we discuss below, we use the optimal Rabi protocols to generate this derived force protocol, to compare it to the optimal force protocol found via pulse engineering.
\begin{figure}
    \centering
    \includegraphics{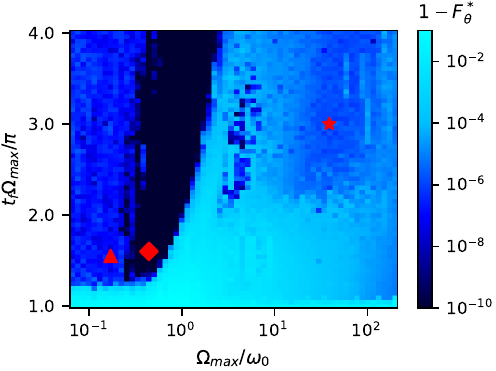}
    \caption{\textbf{Optimal recoil-free state preparation.}
    The optimal infidelity $1-F_\theta^*$ of the optimized protocols, as a function of the maximal Rabi frequency $\Omega_\text{max}$ and the available pulse area $t_f\Omega_\text{max}$, in the absence of dissipation. Three regimes of the maximal Rabi frequency are visually distinguishable: the slow regime with $\Omega_\text{max}/\omega_0 \lesssim  3\times 10^{-1}$, the fast regime for $\Omega_\text{max}/\omega_0 \gtrsim  3\times 10^0$, and the intermediate regime in between. The red markers depict the parameters of the exemplary protocols shown in Fig.~\ref{fig:3}. }
    \label{fig:2}
\end{figure}

\subsection{Optimal protocol}
\label{subsect:optimal-strategies-optimalprot}
We generate an optimal strategy for the Rabi protocol $\Omega(t)$ and the force protocol $f_\text{tw}(t)$, via pulse engineering. Specifically, we parameterize the Hamiltonian by the transformation parameters $\theta=\{\theta_{x,l}, \theta_{y,l}, \theta_{f,l}\}$. We parameterize the Rabi protocol in terms of sine modes~\cite{PhysRevResearch.6.013076}
\begin{align}
    h_{x/y}(t) = \sum_{l=1}^{n_\Omega} \theta_{x/y,l} \sin\left(\pi l \frac{t}{t_f}\right),
    \label{eq:hxy-k}
\end{align}
where $n_\Omega$ is the number of modes of the real and imaginary Rabi protocol. To implement the constraint $|h_x^2(t)+h_y^2(t)| \leq \Omega_\text{max}$ approximately, we check this constraint for a set of times $t_j = j \Delta t$, with $\Delta t = t_f/100$ , in each optimization update, as described below. The force protocol is also parameterized in terms of sine modes
\begin{equation}
    f_\text{tw}^\theta(t) = \sum_{l=1}^{n_f} \theta_{f,l} \sin\left(\pi l \frac{t}{t_f}\right),
    \label{eq:ftw-k}
\end{equation}
where $n_f$ is the number of modes of the force protocol. As described in Sect.~\ref{sec:model}, it is constrained by $v_\text{max}=500 \si{\meter\second^{-1}}$, i.e. $|\partial_t r(t)|~<~v_\text{max}$. To implement this constraint, we evaluate $|\partial_t r(t)|$ for a set of times $t_j = j \Delta t$, similar to the constraint on the Rabi protocol described above. We optimize the transformation parameters $\theta$ using Pulse Engineering via Projection of Response Functions (PEPR)~\cite{heimann_2024-PEPR} with the objective to minimize the infidelity Eq.~\ref{eq:1-F}. For this, we consider the control operators
\begin{align}
    B_x &= \begin{pmatrix}
        0 & e^{i\eta(a+a^\dagger)} \\
        e^{-i\eta(a+a^\dagger)} & 0
    \end{pmatrix},\\
    B_y &= \begin{pmatrix}
        0 & -ie^{i\eta(a+a^\dagger)} \\
        ie^{-i\eta(a+a^\dagger)} & 0
    \end{pmatrix}
\end{align}
for the Rabi protocol. With this we expand the driving Hamiltonian Eq.~\ref{eq:H-dr-1} as
\begin{align}
    H_\text{dr}(t)/\hbar =h_x(t)B_x+h_y(t)B_y.
\end{align}
The third control operator is $B_f=a+a^\dagger$, introduced in the Hamiltonian in Eq.~\ref{eq:Htw}. We proceed as follows. We choose a random time $t_r\in[0,t_f]$ as well as a random control operator $B_j\in\{B_x, B_y, B_f\}$. Then, we compute the susceptibility of the fidelity to the control operator $B_j$
\begin{equation}
    \chi_j(t_r) = \frac{i}{\hbar} \Tr\Big( \rho_* \delta \rho_j(t_r, t_f) \Big),
    \label{eq:chi-final}
\end{equation}
where $\delta \rho_j(t_r)$ is the time-propagated change of the state $[  B_{j},  \rho_\theta(t_r)]$ according to Eq.~\ref{eq:drho}. We then update the transformation parameters as
\begin{equation}
    \theta_{j,l} \rightarrow \theta_{j,l} - \frac{2\alpha_{0,j}}{t_f} \chi_j(t_r)\sin\left(\pi l \frac{t}{t_f}\right).
\end{equation}
Here the hyperparameter $\alpha_{0,j}$ is the phenomenological learning rate. We accept the updated parameters if the corresponding protocols do not violate the imposed constraints, as described above. Otherwise, we discard the correction to the parameters. We iteratively repeat this procedure for a number of $N_\text{it}=10^{5}$ accepted updates.

For the initialization of the optimization algorithm, the transformation parameters of the Rabi frequency $\Omega(t)$ are randomly sampled as $\theta_{x/y,l}\sim \mathcal{N}(0, \Omega_\text{max}/(n_\Omega\sqrt{l}))$. If the resulting protocol violates the imposed constraints, then we discard the parameters and sample another set of transformation parameters $\theta_{x/y,l}$ but with half the value of the standard deviation. We repeat this procedure until a valid protocol is obtained. We initialize the transformation parameters of the force protocol with zero value, i.e. $\theta_{f,l}=0$. In this work we consider $n_\Omega=18$ and $n_f=3$ modes of the protocols. The learning rates, $\alpha_{0,\text{dr}}$ for $B_j \in \{B_x, B_y\}$ and $\alpha_{0,\text{tw}}$ for $B_j = B_f$, are determined heuristically and shown in App.~\ref{app:sec-learning-rates}. This involves initially scanning a range of potential learning rates and subsequently adjusting them based on good performance in generating protocols with low infidelity. We refer to this protocol as the optimized protocol and we refer to the corresponding infidelity as $1-F_\theta$.

\section{High fidelity operations}
\label{sec:high-fidelity-operations}
\begin{figure*}
    \centering
    \includegraphics{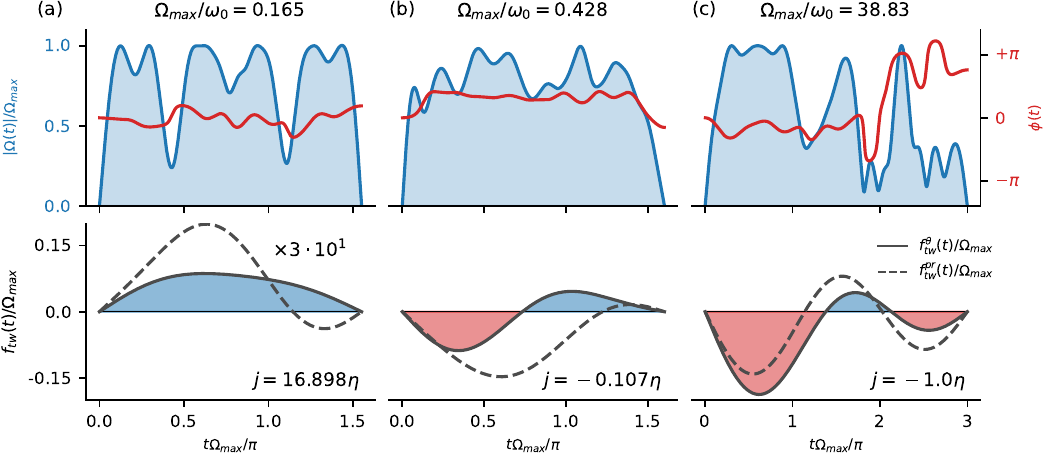}
    \caption{\textbf{Optimal Protocols.}
    We present three exemplary protocols of high-fidelity transformations that yield the highest fidelities observed for the corresponding maximal Rabi frequencies, each chosen from one of the three regimes shown in Fig.~\ref{fig:2}: In panel (a) we show a protocol in the slow regime (red triangle), in panel (b) we show a protocol in the intermediate regime (red diamond), and in panel (c) we show a protocol in the fast regime (red asterisk).  The projected protocol of the restoring potential $f_\text{tw}^{\text{pr}}(t)$ is depicted as a dashed black line. In the fast regime, the optimal protocol of the restoring potential $f_\text{tw}^\theta(t)$ is in good agreement to the projected protocol of the restoring potential $f_\text{tw}^{\text{pr}}(t)$. The restoring potential induces a momentum of $17\hbar k$ in the slow regime, $-0.107\hbar k$ in the intermediate regime, and $-\hbar k$ in the fast regime. In the slow regime, the trap induced dynamics are on the time scales of the transformation time, i.e. $\omega_0 t_f>1$, allowing for deviations from the normalized impulse of $-\eta$, as shown in App.~\ref{app:sec-impulse}.}
    \label{fig:3}
\end{figure*}

To generate optimized protocols, we create an ensemble of optimization trajectories, each parameterized by a set of parameters $\{\theta\}$. We initialize the optimization with a random set of parameters $\{\theta_0\}$. We denote the optimal infidelity of over such an ensemble of optimization trajectories as
\begin{equation}
    1-F_\theta^* = \min_\theta 1-F_\theta.
\end{equation}
In Fig.~\ref{fig:2} we show the optimal infidelity, as a function of the maximal Rabi frequency $\Omega_\text{max}$ and the available pulse area $t_f \Omega_\text{max}/\pi$. There are three main regimes in which optimal strategies emerge. In the slow regime, for $\Omega_\text{max}/\omega_0 \lesssim 3\times 10^{-1}$ and $t_f \Omega_\text{max}/\pi \gtrsim 1.3$, the optimal infidelities are of the order of $10^{-6}$. In the fast regime, for $\Omega_\text{max}/\omega_0 \gtrsim 3\times 10^{0}$ and $t_f \Omega_\text{max}/\pi \gtrsim 1.1$, the optimal infidelities are of the order of $10^{-7}-10^{-4}$. Finally, in the intermediate regime, for $\Omega_\text{max}/\omega_0 \gtrsim 3\times 10^{-1}$ and $t_f \Omega_\text{max}/\pi \gtrsim 1.3$, the optimal infidelities are of the order of $10^{-10}$. Thus, we obtain high-fidelity protocols for a wide range of parameters. The fidelities exceed the currently achieved fidelities, especially in the intermediate regime, which constitutes a globally ideal regime. In App.~\ref{app:sec-labtime} we display the data of Fig.~\ref{fig:2} with the y-axis chosen as $t_f \omega_0/\pi$. Therefore, x- and y-axis utilize the same time scale, resulting in a 'laboratory time' representation. Again, the global optimality of the intermediate regime is transparent. 
We note that the minimal value of $t_f\Omega_\text{max}/\pi$ to achieve low infidelities is larger in the slow regime, for $\Omega_\text{max}/\omega_0 \approx 10^{-1}$, than in the fast regime, for e.g. $\Omega_\text{max}/\omega_0 \gtrsim10^{2}$. The origin of this larger value is the Debye-Waller factor discussed above. For slow Rabi oscillations, the atom is primarily in the motional ground state. As a result, the effective Rabi frequency is reduced by the factor $e^{-\eta^2/2}$.

In Fig.~\ref{fig:3} we show three example protocols, each presenting one of  the three regimes, with $\Omega_\text{max}/\omega_0~=~0.165, 0.428, \text{ and } 38.83$, as shown in Fig.~\ref{fig:2}. Each of these protocols is the optimal protocol for its value of $\Omega_\text{max}/\omega_0$. We present the amplitude $|\Omega(t)|$ and the phase $\phi(t) = \arg \Omega(t)$ of the optimal protocols. In the slow regime, the amplitude and phase of the Rabi protocol are symmetric and anti-symmetric around $t_f/2$, respectively. The protocols at $\Omega_\text{max}/\omega_0=0.165$ and $\Omega_\text{max}/\omega_0=0.428$ are optimal for $t_f\Omega_\text{max}/\pi\approx 1.5$, and thus not significantly larger than the naive value of $t_f\Omega_\text{max}/\pi=1$, or the Debye-Waller corrected estimate of $t_f\Omega_\text{max}/\pi=e^{\eta^2/2}$. As a result, the optimal protocols feature an amplitude $|\Omega(t)|$ that is close to $\Omega_\text{max}$ for most of the time interval. The third protocol has an available pulse area of $t_f\Omega_\text{max}/\pi\approx3$ and therefore the ratio of the pulse area to the available pulse area $\Theta/(t_f\Omega_\text{max})$ is reduced, compared to the other two cases. 

Additionally, the optimized protocols $f_\text{tw}^\theta(t)$ are shown as a function of time. We note that the optimal force protocols are dominated by a different frequency in the three regimes, by $l=1$, $l=2$, and $l=3$, for the slow, intermediate, and fast regime, respectively. For comparison, we show projected force protocols, which are based on the optimal Rabi pulse that we obtained via optimization. We observe that the projected protocol approaches the optimal force protocol in the fast regime, see the lower panel of Fig.~\ref{fig:3} (c). We note that in the fast regime, the protocols have a normalized impulse of $j=-\eta$. This is consistent with the observation that in the fast regime, the protocol of the restoring potential is the main degree of control to mitigate photon induced recoil. We note that we have also performed optimization without utilizing the restoring force. The resulting protocols do not implement a mitigation of the photon induced recoil in the fast regime. Therefore, the protocol of the restoring potential $f_\text{tw}(t)$ is crucial for high-fidelity operations in the fast regime and leads to an improvement of the infidelity of several orders of magnitude. In the slow regime, the normalized impulse $j$ and the form of the restoring potential differ  from the recoil-compensated Rabi protocol. In App.~\ref{app:sec-impulse} we show the normalized impulse for all values of $\Omega_\text{max}$.

\begin{figure}
    \centering
    \includegraphics{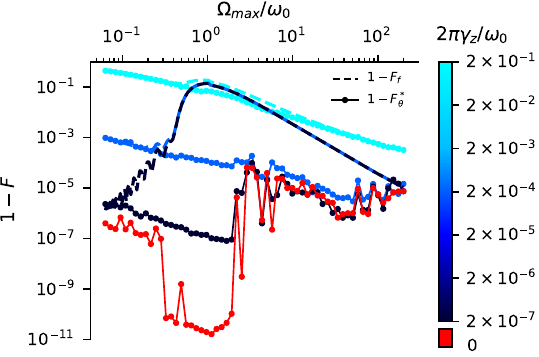}
    \caption{\textbf{Dissipation.}
    The infidelity $1-F$, as a function of the Rabi frequency $\Omega_\text{max}/\omega_0$ for a fixed available pulse area of $t_f\Omega_\text{max}/\pi = 3.5$, in the presence of dissipation for different values of $\gamma_z$ in units of $\omega_0/2\pi$. We show the infidelity of the recoil-compensated Rabi protocol $1-F_f$ (dashed lines), the optimal infidelity $1-F_\theta^*$ for varying dissipation rates (solid lines with circular markers), and the optimal infidelity $1-F_\theta^*$ in absence of dissipation (solid red line with circular markers). Dissipation imposes a lower bound on the infidelity. Both in the slow and in the fast regime, the optimal infidelity converges to the infidelity of the recoil-compensated Rabi protocol.}
    \label{fig:4}
\end{figure}

{\it Dissipation.} In Fig.~\ref{fig:4} we show the infidelity of the recoil-compensated Rabi protocol $1-F_f$ as well as the optimal infidelity $1-F_\theta^*$ in presence of dissipation. As described above, we consider dephasing with a rate $\gamma_z$ as an example. We keep the available pulse area fixed at $t_f\Omega_\text{max} = 3.5\pi$. Introducing dissipation leads to a lower bound to the infidelity, which is of the order of $\gamma_zt_f$, or, for fixed pulse area, of the order of $\gamma_z/\Omega_\text{max}$. However, the optimal protocols display high-fidelity transformations for a wide range of maximal Rabi frequencies. As shown in Fig.~\ref{fig:4}, the recoil compensated Rabi protocol generates low infidelities for both the slow and the fast regime, but generates high infidelities around $\Omega_\text{max}/\omega_0\approx 1$. For small dephasing rates, such as $\gamma_z = 2\times10^{-7}$ in units of $\omega_0/2\pi$, the optimal protocols improve on the infidelities of the recoil-compensated Rabi protocol by up to six orders of magnitude. For $\gamma_z = 2\times 10^{-4}$ in units of $\omega_0/2\pi$, an improvement of 3 orders of magnitude is achieved. We note that for both the slow and fast regime, the optimal protocols approach the performance of the recoil-compensated Rabi protocol, but generate significant improvement in the intermediate regime.

\section{Conclusion}
\label{sec:conclusion}
In conclusion, we have demonstrated recoil-free state preparation of an atom in an optical tweezer. We propose control protocols that utilize both the phase and amplitude of the control laser, as well as the spatial displacement of the tweezer position. Specifically, we consider a model of two states that can be driven by a single-photon Rabi pulse.  We consider finite harmonic confinement of the atom. Thus, the photon recoil of the Rabi pulse drives the spatial motion of the atom in the tweezer, with a coupling strength given by the Lamb-Dicke parameter. This motion takes the system out of the logical subspace, we therefore generate optimal protocols via pulse engineering to suppress the motion, with our central example of creating a $\pi$-pulse of $\ket{0}$ to $\ket{1}$ with maximal fidelity. We use the maximal Rabi frequency and the maximal pulse area as a parametrization of the system.  We show that the ratio of the maximal Rabi frequency and the trap frequency is the key quantity that determines the overall fidelity regime, if sufficient maximal pulse area is provided. We find three regimes, which we refer to as slow, intermediate and fast, with infidelities well below current technological standards. In the fast regime the numerically optimized protocol is comparable to a recoil-compensated Rabi protocol that we propose. For intermediate regimes we provide a numerical optimal protocol based on pulse engineering. With these results we point out both a general approach of optimal pulse engineering for Lamb-Dicke coupled qubits, as well as a concrete realization that is applicable to current atomic tweezer systems.

\begin{acknowledgments}
This work is funded by the Deutsche Forschungsgemeinschaft (DFG, German Research Foundation) - SFB-925 - project 170620586 and the Cluster of Excellence 'Advanced Imaging of Matter' (EXC 2056) project 390715994. The project is co-financed by ERDF of the European Union and by ’Fonds of the Hamburg Ministry of Science, Research, Equalities and Districts (BWFGB)’.
\end{acknowledgments} 

\bibliography{main.bib}

\appendix

\section{Derivation of Eq.~\ref{eq:ev-expval}}
\label{app:sec-derivation-exp}
We propose the force protocol 
\begin{equation}
    f^{h}_\text{tw}(t) = \eta\frac{h(t)}{2}\langle \sin(\varphi)\sigma_x(t) + \cos(\varphi)\sigma_y(t) \rangle
     \label{eq:app-ev-expval}
\end{equation}
to compensate recoil at first order in $\eta$. The contained expectation value is evaluated at 0-th order in $\eta$. Eq.~\ref{eq:app-ev-expval} is derived expanding the driving Hamiltonian Eq.~\ref{eq:H-dr-1} to first order in $\eta$ \begin{equation}
\begin{split}
    H_\text{dr}/\hbar &\approx \frac{h(t)}{2}(\cos(\varphi)\sigma_x-\sin(\varphi)\sigma_y) \\&- \eta\frac{h(t)}{2}(\sin(\varphi)\sigma_x + \cos(\varphi)\sigma_y)(a+a^\dagger).
\end{split}
\end{equation} 
The corresponding unitary in 0-th order in $\eta$ is
\begin{equation}
    \begin{split}
    U_\varphi(t) =& \cos\left(\int_0^{t} d\tau \frac{h(\tau)}{2}\right)\sigma_0 + \\
    &-i\sin\left(\int_0^{t} d\tau \frac{h(\tau)}{2}\right)(\cos(\varphi)\sigma_x-\sin(\varphi)\sigma_y).
    \end{split}
\end{equation}
To calculate the expectation value in Eq.~\ref{eq:app-ev-expval}, we consider the state $\ket{\psi(t)}=U_\varphi(t)\ket{0}$, resulting in Eq.~\ref{eq:ev-expval}: 
\begin{equation}
    \langle \sin(\varphi)\sigma_x(t) + \cos(\varphi)\sigma_y(t) \rangle = -\sin\left(\int_0^{t} d\tau h(\tau)\right).
\end{equation}

\section{Properties of the restoring potential protocol}
\label{app:sec-impulse}
\begin{figure}
    \centering
    \includegraphics{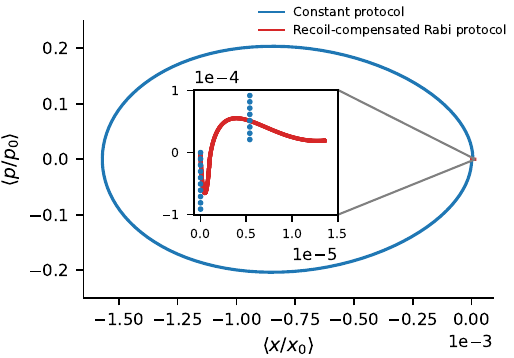}
    \caption{\textbf{Phase space.} 
    Phase space dynamics for $\Omega_\text{max}/\omega_0=200$. We show the resulting phase space of the constant restoring force protocol $f_\text{tw}(t)~=~-\eta/t_f$ (blue) as well as the resulting phase space of the recoil-compensated Rabi protocol Eq.~\ref{eq:ftw} (red). The maximal displacement is reduced by two orders of magnitude in the case of the recoil-compensated Rabi protocol compared to the constant protocol and the maximal momentum is reduced by three orders of magnitude.
    }
    \label{app:fig_phasespace}
\end{figure}

In the following, we  discuss the phase space dynamics of two different protocols and the normalized impulse resulting obtained from the optimized protocols. We have demonstrated, that in the fast regime, a restoring potential protocol that satisfies Eq.~\ref{eq:j} reduces motional excitations in the state preparation transformation. While achieving qualitatively similar fidelities, the phase space dynamics vary for different protocols. In Fig.~\ref{app:fig_phasespace} we show the phase space dynamics in the fast regime for two restoring potential protocols. The first protocol has a constant amplitude $f_\text{tw}(t)~=~-\eta/t_f$ and the second protocol is the recoil-compensated Rabi protocol Eq.~\ref{eq:ftw}. The latter results in dynamics reduced by several orders of magnitude compared to the former, with a reduction of three orders of magnitude in the momentum and two orders of magnitude in the displacement.

\begin{figure}
    \centering
    \includegraphics{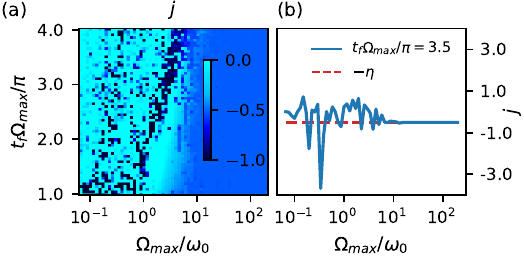}
    \caption{\textbf{Normalized impulse.}
    (a) The normalized impulse j resulting from the optimized protocols, as a function of the maximal Rabi frequency $\Omega_\text{max}$ and the available pulse area $t_f\Omega_\text{max}$. (b) The normalized impulse for a fixed value of the available pulse area $t_f\Omega_\text{max}/\pi = 3.5$ (blue line) and $j=-\eta$ (red line). In the fast case, the normalized impulse converges against $-\eta$.}
    \label{app:fig_impulse}
\end{figure}

In Fig.~\ref{app:fig_impulse} we show the normalized impulse 
\begin{equation}
    j = \int_0^{t_f} f_\text{tw}^\theta(t).
\end{equation}
from the optimized protocols, as a function of $\Omega_\text{max}$ and $t_f\Omega_\text{max}$. We find that in the fast regime, the normalized impulse is $j=-\eta$, so that Eq.~\ref{eq:hk} is fulfilled.

\section{Learning rates}
\label{app:sec-learning-rates}
In this section, we show the learning rates $\alpha_{0,\text{dr}}$ for $B_j \in \{B_x, B_y\}$ and $\alpha_{0,\text{tw}}$ for $B_j = B_f$ used to generate the optimized protocols in Fig.~\ref{fig:2}, which are found heuristically. The initial learning rates $\hat{\alpha}_{0,\text{dr}}$ and $\hat{\alpha}_{0,\text{tw}}$ are found by performing optimization processes for the Rabi frequencies $\Omega_\text{max}/\omega_0$ for a coarser time grid $t_\text{test}\Omega_\text{max} \in \{\pi, 1.5\pi,\cdots, 4\pi\}$. In the test optimization processes, a scan over  $\hat{\alpha}_{0,\text{dr}},\hat{\alpha}_{0,\text{tw}} \in \{0.01,0.21,\cdots,2.01\}$ is performed, identifying the optimal learning rates. These are then used for the whole span of transformation times $t_f\Omega_\text{max}$ by assigning each optimization process the learning rates found for the closest $t_\text{test}\Omega_\text{max}$ respectively. In general, choosing learning rates that are too large leads to unstable performance, preventing the algorithm from finding high-fidelity protocols.

\begin{figure}
    \centering
    \includegraphics{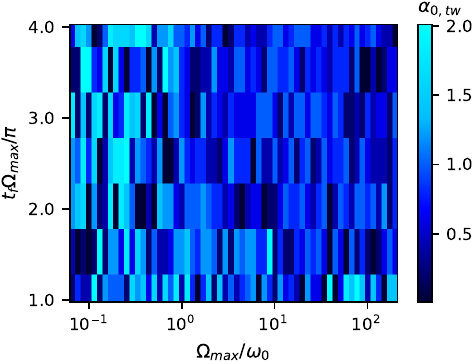}
    \caption{\textbf{Learning rates $\alpha_\text{0,tw}$ for $B_j = B_f$.}
    Learning rates $\alpha_\text{0,tw}$ as a function of $\Omega_\text{max}$ and the transformation time, found by scanning over a range of learning rates. The learning rates are small in the fast regime, i.e. for $\Omega_\text{max}\gtrsim \omega_0$.}
    \label{app:fig_alphaForce}
\end{figure}

$\alpha_{0,\text{tw}}$, which is identical to $\hat{\alpha}_{0,\text{tw}}$ in our case, is shown in Fig.~\ref{app:fig_alphaForce} as a function of the maximal Rabi frequency and the transformation time. For the regime where $\Omega_\text{max}\gtrsim \omega_0$, we find smaller values than in the slow regime, i.e.  $\Omega_\text{max}< \omega_0$. This is also the region where the restoring force is the main degree of freedom to generate high-fidelity protocols and reduce photon induced recoil. 

\begin{figure}
    \centering
    \includegraphics{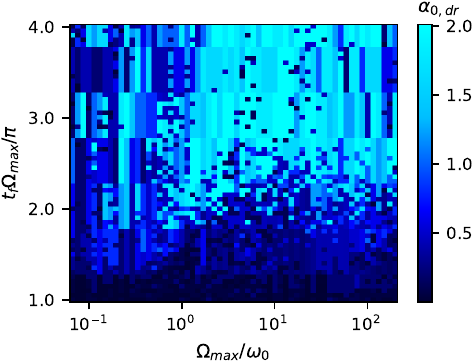}
    \caption{\textbf{Learning rates $\alpha_\text{0,dr}$ for $B_j \in \{B_x,B_y\}$.} 
 Learning rates $\alpha_\text{0,dr}$ as a function of $\Omega_\text{max}$ and the transformation time, found by scanning over a range of learning rates and further reduced based on performance. We find small learning rates for transformation times $t_f\Omega_\text{max} \lesssim 1.7\pi$, as well as in the slow regime where $\Omega_\text{max}<\omega_0$.}
    \label{app:fig_alphaRabi}
\end{figure}

$\alpha_{0,\text{dr}}$ is shown in Fig.~\ref{app:fig_alphaRabi}, as a function of the maximal Rabi frequency and the transformation time as well. In this case the values of the learning rates $\hat{\alpha}_{0,\text{dr}}$ are adjusted further. This is done based on performance observation and consequently reducing $\alpha_{0,\text{dr}}$ by division by powers of two, until convergence is reached. For transformation times $t_f\Omega_\text{max} \lesssim 1.7\pi$ we find similar learning rates across the entire range of maximal Rabi frequencies $\Omega_\text{max}$, with values not exceeding $\alpha_{0,\text{dr}}\approx 1$. For longer transformation times, the learning rates take on larger values in the fast regime, but overall remain small in the slow regime, i.e. for $\Omega_\text{max}/\omega_0\lesssim 1$. 

\section{Laboratory time}
\label{app:sec-labtime}
\begin{figure}
    \centering
    \includegraphics{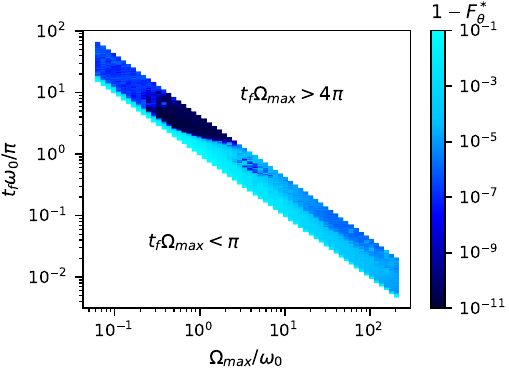}
    \caption{\textbf{Optimal state preparation in laboratory time.}
    We show the optimal infidelity $1-F_\theta^*$ presented in Fig.~\ref{fig:2} as a function of the maximal Rabi frequency $\Omega_\text{max}$ in units of the trap frequency $\omega_0$ and the laboratory time $t_f\omega_0$ in units of $\pi$, with both axes being displayed on a logarithmic scale. At $t_f\omega_0\gtrsim 2\pi$ a plateau emerges, where high-fidelity implementations are found for a range of maximal Rabi frequencies. This suggests the choice of such a laboratory time in the intermediate regime.}
    \label{app:fig_labtime}
\end{figure}
In this section, we discuss the laboratory time of the optimal state preparation. In Fig.~\ref{app:fig_labtime} we display the optimal infidelity $1-F_\theta^*$ of Fig.~\ref{fig:2} as a function of the maximal Rabi frequency $\Omega_\text{max}$ and the laboratory time $t_f\omega_0$. Note that both the x-axis and the y-axis are on a logarithmic scale. The data-free triangle on the lower left corresponds to transformation times $t_f\Omega_\text{max} < \pi$ and the upper right triangle corresponds to transformation times $t_f\Omega_\text{max} > 4\pi$. Again, the three regimes are visually distinguishable. For a laboratory time of $t_f\omega_0\approx 2\pi - 4\pi$, we observe high fidelities over a wide range of Rabi frequencies. This indicates that selecting a laboratory time within this range is beneficial and well-suited when operating in the intermediate regime, regardless of the exact value of $\Omega_\text{max}$.

\section{Stroboscopic Infidelity}
\label{app:sec-stroboscopic-infidelity}
\begin{figure}
    \centering
    \includegraphics[width=7.5cm]{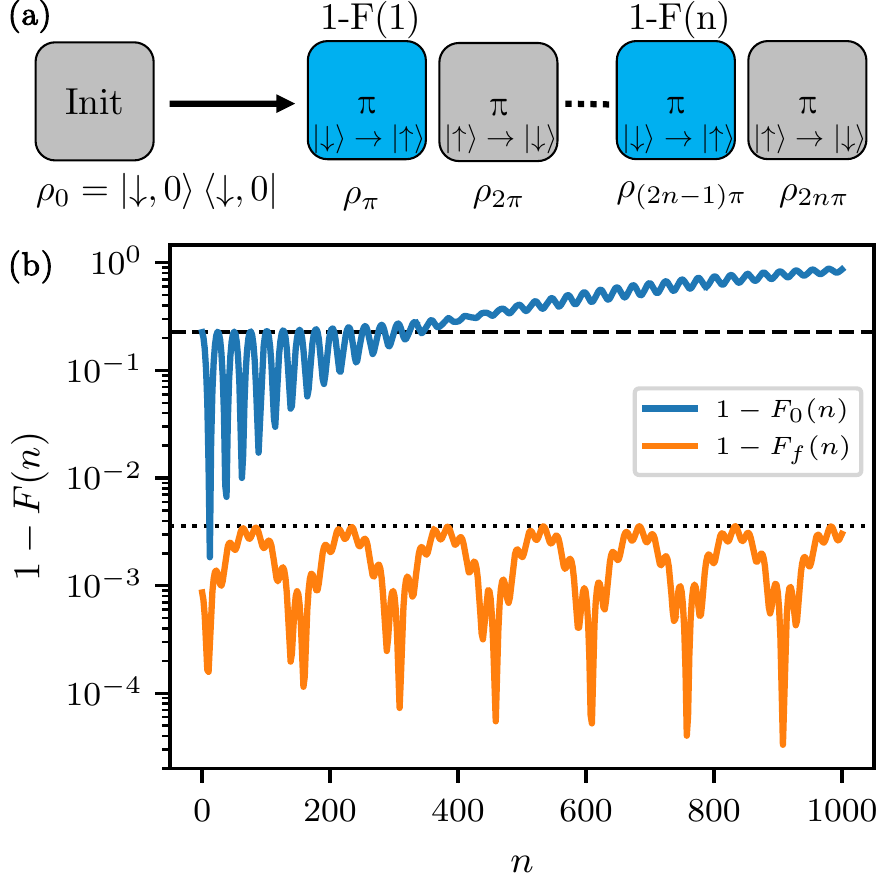}
    \caption{\textbf{Stroboscopic infidelity.}
    (a) Illustration of the stroboscopic pulse scheme. Starting from the initial state $\rho_0$, we apply a sequence of $\pi$-pulses and interrogate the infidelity $1-F(n)$ after an $(2n-1)$ pulses. (b) The infidelity $1-F(n)$ of the state preparation, as a function of the number of repetitions $n$ for $\Omega_\text{max}/\omega_0 = 20$, i.e. in the fast regime. The infidelity based on the Rabi protocol $1-F_0(n)$ is depicted as the blue line and the infidelity based on the recoil-compensated Rabi protocol $1-F_f(n)$ is depicted as the orange line. The infidelity $1-F_0(n)$ obtains local minima at $(2m-1)\Omega_\text{max}/(2\omega_0)$ for $m > 0$. For $n>350$ repetitions, the infidelity using the Rabi protocol exceeds the initial value $1-F_0(n) > 1-F_0(1)$, depicted by the black dashed horizontal line, as the system heats up. However, the infidelity using the recoil-compensated Rabi protocol obtains an upper bound of $1-F_f(n) < 3.5\times 10^{-3}$ for all $n$, depicted by the black dotted horizontal line, demonstrating that motional excitations are confined.}
    \label{app:fig_strobo}
\end{figure}

In this section we show that the recoil-compensated Rabi protocol confines excitations of the internal states over many repetitions in the fast regime. We consider the initial state $\rho_0$. Next, we perform the state preparation transformation based on the recoil-compensated Rabi protocol. Subsequently, we apply a second transformation but with the opposite restoring potential $f_\text{tw}^{-\sin}(t) = -f_\text{tw}^{\sin}(t)$. We repeat this procedure and analyze the infidelity $1-F(n)$ of the $(2n-1)$-th transformation, as illustrated in Fig.~\ref{app:fig_strobo}~(a). We choose $\Omega_\text{max}/\omega_0 = 20$. Fig.~\ref{app:fig_strobo}~(b) shows the infidelity $1-F_0(n)$ and $1-F_f(n)$ over a sequence of $2n$ transformations using the Rabi protocol and the recoil-compensated Rabi protocol, respectively. In the case of the Rabi protocol, $1-F_0(n)$ oscillates between $10^{-1}$ and $10^{-3}$ with a phase of $n\pi \omega_0/\Omega_\text{max}$ and the infidelity obeys local minima $1-F_0(n^*)$ where $n^*=(2m-1)\Omega_\text{max}/(2\omega_0)$ for $m > 0$. For large $n$, the values of the minima $1-F_0(n^*)$ increase monotonically. For $n>350$ repetitions, the infidelity exceeds the value of the initial infidelity $1-F_0(n) > 1-F_0(1)$. Using the recoil-compensated Rabi protocol, we find $1-F_f(1) = 9 \times 10^{-4}$ for this set of parameters. In contrast to the Rabi protocol, the infidelity obtains an upper bound of $1-F(n) <  3.5 \times 10^{-3}$ for all $n$. Therefore, the recoil-compensated Rabi protocol allows to confine excitations of the motional states over many repetitions.

\end{document}